# Electrophoretic-like gating used to control metal-insulator transitions in electronically phase separated manganite wires

*Hangwen Guo,[1,2] Joo H. Noh,[3,4] Shuai Dong,[1,2,5] Philip D. Rack,[3,4] Zheng Gai,[4] Xiaoshan Xu,[1]*

*Elbio Dagotto,[1,2] Jian Shen,*[6,2] and T. Zac Ward*[1]*

[1]Materials Science and Technology Division, Oak Ridge National Laboratory, Oak Ridge, TN 37831, USA, [2] Department of Physics & Astronomy, University of Tennessee, Knoxville, TN 37996, USA, [3] Materials Science & Engineering, University of Tennessee, Knoxville, TN 37996, USA, [4] Center for Nanophase Materials Sciences, Oak Ridge National Laboratory, Oak Ridge, TN 37831, USA, [5] Department of Physics, Southeast University, Nanjing 211189, China, [6] Department of Physics, Fudan University, Shanghai, 200433, China

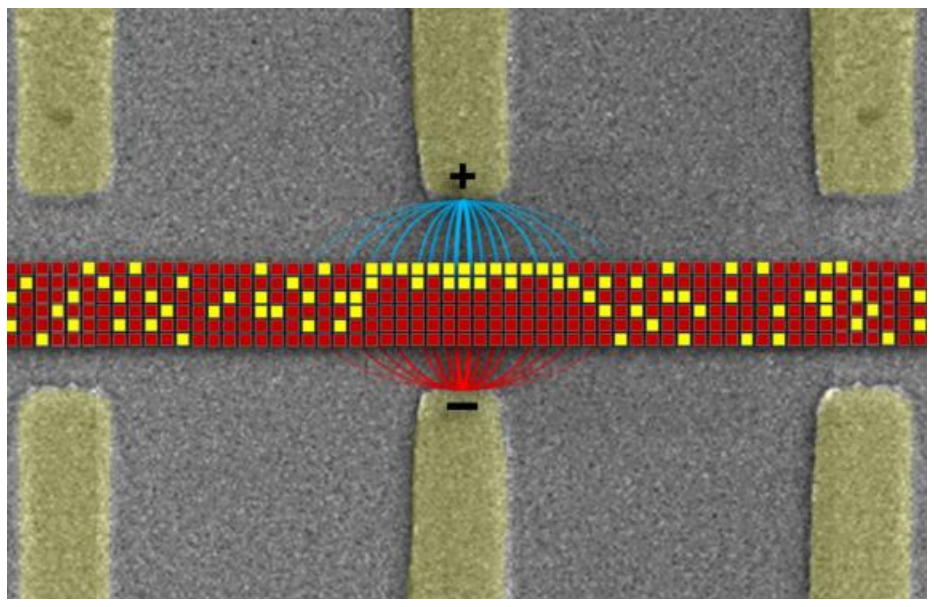

*Electronically phase separated manganite wires are found to exhibit controllable metal-insulator transitions under local electric fields. The switching characteristics are shown to be fully reversible, polarity independent, and highly resistant to thermal breakdown caused by repeated cycling. It is further demonstrated that multiple discrete resistive states can be accessed in a single wire. The results conform to a phenomenological model in which the inherent nanoscale insulating and metallic domains are rearranged through electrophoretic-like processes to open and close percolation channels.*

Resistive switching is observed across many different material classes and can be driven by widely different mechanisms ranging from valence change and electrostatics to molecular fluid flow and nanomechanical processes.[1] Complex materials are very promising candidates for discovery of new switching processes since their strongly correlated electronic properties often create a finely balanced system that can be drastically modified with small changes to the underlying order parameters. These materials are of high interest for next generation applications since they possess many compelling properties such as high Tc superconductivity, multiferroicity, and colossal magnetoresistance. In many cases, these behaviors are accompanied by coexisting electronic phases of vastly different resistive and magnetic character at the nanoscale.[2–6] Spatially manipulating these phases while preserving the novel macroscopic behaviors offers the potential for creating new types of electronic devices with new degrees of functionality.

Resistive switching has been achieved in several of these systems using traditional carrier doping and by inducing electroresistive phase transitions in the material. In the case of carrier manipulation, a gate electrode is directly applied to the surface of the studied material and an electric field applied to the gate acts to change the material's access to electrons. This switching behavior has a wide, active temperature range, shows a relatively linear change in resistance with applied bias, and has a strong bias polarity dependence.[7–10] In contrast, switching devices based on electronic phase transitions rely on large voltages that are directly applied along the probe direction in the device, thereby inducing a first order transition in the electronic states. These devices are characterized by a wide, active temperature range, require switching voltages that are strongly temperature dependent, and show static or large hysteresis in switching behavior arising from the free energy attributed to a first order transition.[1,11–17]

We demonstrate that lateral gate electrodes can be used on electronically phase separated manganite nanowires to induce dynamic metal-insulator transitions with novel switching characteristics that do not conform to any mechanisms previously reported. To explain the observed behavior, we introduce a phenomenological model in which conduction channels are formed within the wire by manipulating the positions of the inherent metallic phase domains. The unique means of switching and device structure allow non-static switching, reduced risk of thermal breakdown on repeated switching, have no subthreshold current leakage, provide bias-polarity independence, and can be used in a geometry that gives access to multiple discrete resistive states along a single wire. The characteristics observed offer new possibilities for accessing low power switching, creating reconfigurable interconnects, and open a new mechanism for application in novel electronic devices.

In electronically phase separated materials, regions with very different resistive and magnetic properties can coexist on length scales ranging from microns to nanometers.[4] While there is a great deal of debate on what gives rise to this phenomenon, it is well accepted that the strongly correlated spin-charge-lattice-orbital order parameters are of central importance in how these phases seed and coexist.[4,16,18] Even small variations to the underlying energetics can have dramatic effects on phase behavior and can lead to colossal changes in character. Most proposed device applications in these systems are based on controlling the macroscopic behaviors by applying global field tuning such as substrate strain, electric and magnetic fields, or thermal manipulation.[5,16,19–22] Recent work on materials confined to length scales similar to that of the phase domains has shown that it is possible to create structures in which transport is dominated by a relatively few domains.[3,23–25] The combination of confined structures with local field tuning may then be expected to allow access to another layer of tunability in these materials. In this work, we describe an example of this approach where inherent metallic and insulating electronic phases are manipulated in a single crystal manganite wire through the application of local electric fields.

The ideal prototype material to investigate locally controllable phase separated circuitry needs to have domain sizes that are accessible with standard confinement techniques and must have a region known to possess energetically balanced electronic phases of significantly different resistive properties. For these reasons, we selected the colossal magnetoresistive manganite $[La_{1-x}Pr_x]_{5/8}Ca_{3/8}MnO_3$ (x=0.3) (LPCMO) which has charge-ordered insulator (COI) and ferromagnetic metal (FMM) domains coexisting near the metal-insulator transition and possesses a fluid phase separated state (or strain liquid phase) where the coexisting phases strongly interact to maintain local energetic balance.[14,26–30] This fluid phase separated state is defined by the thermal region in which the FMM and COI phases have a similar free energy.[28,29,31] In this regime, the competing phases are not pinned and can be described as electronically soft matter, where even slight changes to the energetic landscape in which the phases reside can drive domain orientations.[16,29,31] A 50 nm thick single crystal manganite film on $SrTiO_3$ (001) and gate electrodes were patterned using both photo and electron beam lithography to create 4 pairs of 40 μm x 40 μm pads connected by 20 μm long wires of 400 nm, 700 nm, 1 μm, and 2 μm widths. (See supporting Information for fabrication details) Figure 1a and 1b show an example of the experimental device design where a manganite wire is set in a 2-probe geometry with 3 sets of evenly spaced freestanding lateral gates along the wire length with 1 μm vacuum gaps. The resistance is monitored along the wires with a 50 nA constant current source. Bi-polar voltages are applied to the Au gate electrodes and induce an electric field through the vacuum gaps and patterned LPCMO wire.

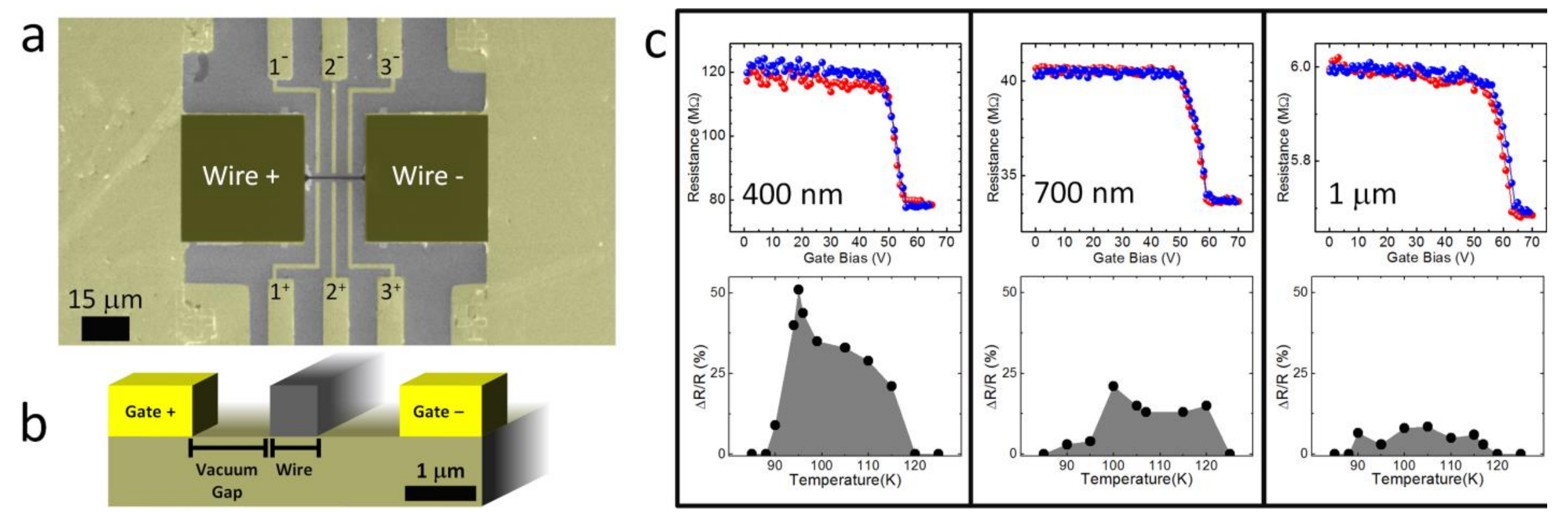

**Figure 1.** Device geometry and properties of $[La_{1-x}Pr_x]_{5/8}Ca_{3/8}MnO_3$ wires under electric field. **(**a)SEM image of device configuration showing 3 sets of lateral Au gates placed along manganite wire, color overlay for clarity. (b)Cross sectional drawing of LPCMO wire near a set of Au gates where scale bar is to scale in-plane direction only. (c)Top: Resistance versus gate voltage curves under 3 T magnetic field for 400 nm, 700 nm, and 1 μm wires with electric field applied through central gate electrodes. Each curve shown is taken at the individual peak resistance switching temperature. Bottom: Maximum relative resistance change across a range of temperatures for each wire.

Figure 1c compares the behavior of 400 nm, 700 nm, and 1 μm wide LPCMO wires when applying an electric field across the central gate. In each case, there is a sharp change in resistance above a critical field without obvious hysteresis or residual transition effects when the electric field is removed from the gate. The relative change in resistance is strongly tied to wire width, where narrower wires have the highest percent change in resistance which means that these devices have excellent scaling properties. This is not surprising since the smaller wires have fewer possible percolation paths.[23,25] While outside the scope of this work, we note that the relative resistance change could be greatly enhanced by using a single gate on a shorter wire where only the local resistance change would dominate. Of particular interest, is that switching occurs in a limited temperature range of ~ 85 – 120K but is never observed outside of this temperature window in any wire width. This behavior is unlike any previously observed in complex oxide devices and suggests that a new mechanism is driving these resistance changes.[1,11–15]

Figure 2a shows the 400 nm wire's resistance as its central gate's voltage is cycled from 0 V to 60 V to -60 V to 0 V. For consistency in determining the switching voltage, we use the peak in the slope of resistance as a function of applied voltage. The resistance switch occurs at the same voltage regardless of ramping direction or polarity. The resistance is also flat above and below the critical switching voltage. Figure 2b presents an example of the simulated electric field intensity applied across a 400 nm wire. To model the contour of electric field intensity, two-dimensional cross-sectional dielectric matrices are constructed with the same geometry as the experimental devices. The electric field is calculated for the entire device cross-section including the manganite wire, $SrTiO_3$ substrate, vacuum gap, and Au contacts. The dielectric values of the Au gates are set to a sufficiently high value of 10,000. The phase separated manganite wires are assigned a dielectric value of 3000 while vacuum and $SrTiO_3$ substrate dielectric constants are taken as the standard values: 1 and 1000, respectively.[13,32,33] The experimentally observed switching potentials are then applied to the Au gates and the local electric potential and field is solved using a differential equation of Gauss' law ($\nabla \cdot D = 0$, where $D$ is the electric displacement field) across the matrices. This was solved for multiple temperatures on each of the three wires. Figure 2c gives the maximum electric field values within the wires at each temperature when switching occurred. (See Supporting Information for more details) Two key observations from this are that the switching field is not temperature dependent and that the electric field within the wire when the resistance drop occurs appears to be independent of wire width. These observations are in sharp contrast to previously observed resistive switching behaviors in other devices where the driving mechanism is attributed to either carrier doping, band bending, or phase transitions.[7,8,13,34]

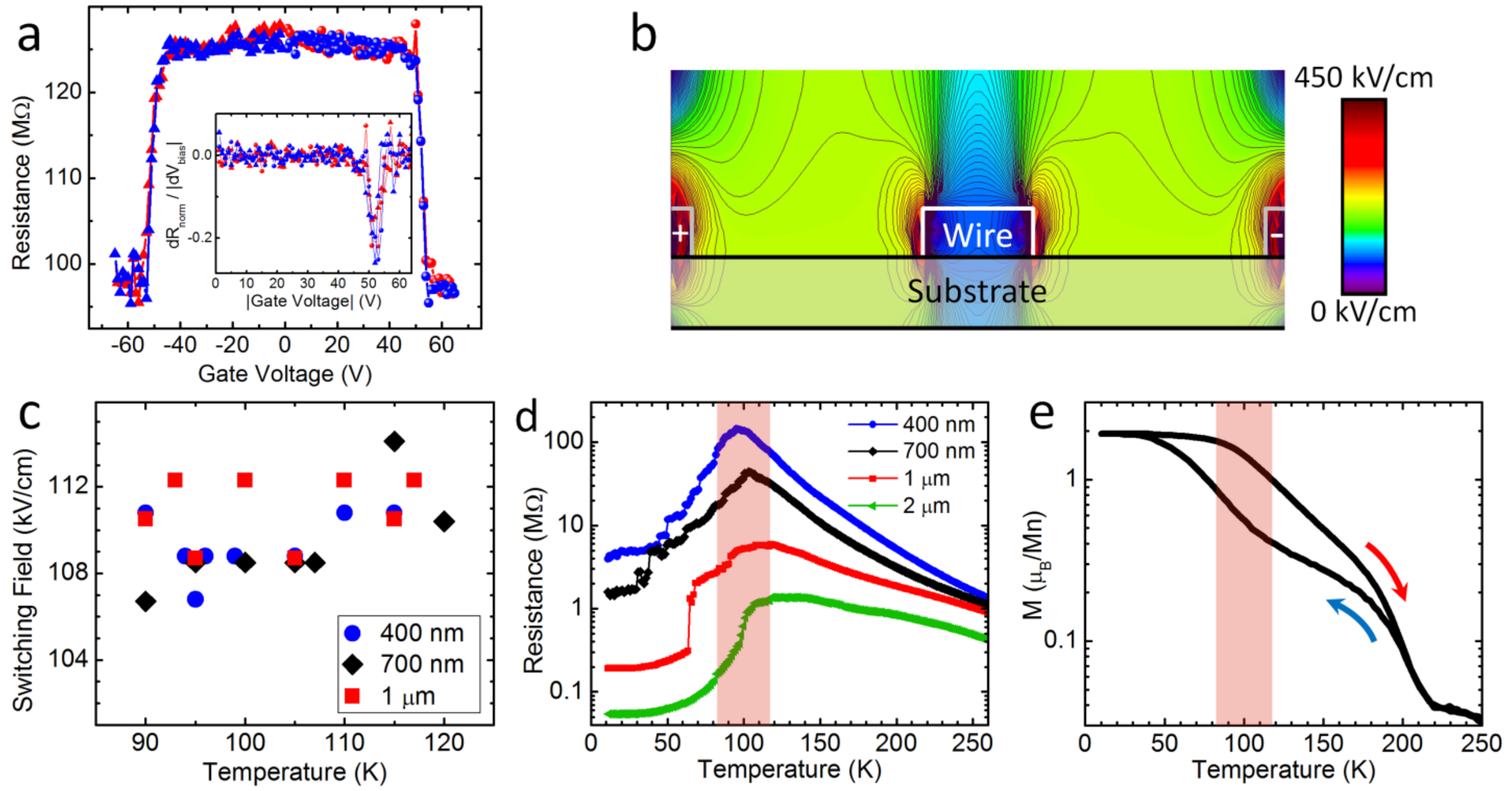

**Figure 2.** Bias cycling response and temperature dependence of switching fields, resistance, and magnetization. (a)Example of forward(circle) and reverse(triangle) field effects on increasing(red) and decreasing(blue) gate voltage show no hysteresis and symmetric response to bias polarity. Taken on the 400 nm wire using central gate at 99K under a 3 T magnetic field at a 1V/s rate. (*Inset*) Peak in $\Delta R/\Delta V_{gate}$ occurs at the same electric field regardless of polarity or ramping direction. This peak is taken as the switching voltage. (b)Example of a 2D electric field map used to find maximum field within the wires at the recorded switching voltages. (c)Switching electric field values for wires across active temperature range shows no dependence on temperature or width. (d)Resistance vs temperature on cooling 5K/min under a 3 T magnetic field and 0V gate voltage for 20 μm long wires of varying width etched from the same 50 nm thick single crystal thin film. (e)Field cooling and warming magnetization vs temperature plots under a 0.1 T magnetic field for a parent LPCMO film before lithographic processing. Highlighted regions in (d) and (e) show the temperature window between onset of ferromagnetic metal phase seeding and blocking temperature where fluid phase separation exists.

Carrier doping is a common mechanism applied to changing a material's resistance by injecting electrons across an interface into the device. Properties that go along with this mechanism are a subthreshold region in which the resistive gate is being opened and can be characterized by a $\Delta R/\Delta V_{gate}$ having a non-zero value for $V_{gate} > 0$, bias polarity dependence, and a broad active temperature range.[32,33] Carrier doping is ruled out as a possible mechanism in the laterally gated devices, since there is no evidence of a similar subthreshold region upon voltage application, no electroresistive sign dependence on bias polarity, and the active temperature range is confined to the fluid phase separated temperature range.[7,8] Further, piezoelectric influences induced at the $SrTiO_3$ interface can also be ruled out because only a small volume of the substrate is affected by electric field and this is clamped by the underlying substrate. This leaves the possibility that the resistance changes can be attributed to electric field induced phase transitions; however, there are several reasons why this seems unlikely. A charge ordered insulator to ferromagnetic metal phase transition is of first order; this means that there is a free energy component that must be accounted which accompanies the change in metallic volume fraction.[16] This fact makes the switching fields strongly temperature dependent, allows switching across a wide temperature range, and gives rise to a relatively static switching behavior with large hysteresis on electric field cycling.[1,11–15] While it is impossible to rule out small contributions from these mechanisms in the present study, the vastly different switching characteristics make it clear that the dominant process involved in creating the large abrupt opening of the conducting percolation path in the side gated devices is not due to any previously reported mechanisms.

To understand why these devices act as they do, the resistive and magnetic properties of the un-biased material must be considered. Figure 2d shows the resistance as a function of temperature at zero gate voltage. All wires except the widest present sharp jumps in resistance along the metal-insulator transition which indicates that transport is being dominated by a small number of phase domains transitioning from insulating to metallic states.[23,25,30,35] The magnetization curve of the un-patterned manganite film in figure 2e shows the ferromagnetic metal phase onset below 120 K on cooling. The temperature at which the plateau in the warming curve of manganetization begins to sharply decrease, ~80K, can be taken as the low temperature limit of the fluid phase separated state.[19] The transport and magnetic measurements when taken together are consistent with the material possessing active coexistence of electronic phases balanced in the fluid phase separated state.[19,20] This region is characterized by the presence of coexisting metal and insulating phases that are energetically balanced and unlocked from the low temperature glass phase and is highlighted in figures 2d and 2e between 80K and 120K which is also the temperature range in which switching is observed.[14,26–30]

Based on these observations, a phenomenological electrophoresis model using a two-dimensional ($L_x$x$L_y$) lattice is adopted to simulate the resistance in the manganite nanowires across a range of temperatures and applied electric fields. A matrix is constructed from metal and insulating sites where each phase type is assigned a charge constant, $Q_M$ for the metallic site and $Q_I$ for the insulating site which does not break the charge neutrality of the material. The Hamiltonian is given as: $H = \sum_i Q_i V_i + A \sum_{<i,j>} Q_i Q_j$. The first term represents the on-site potential energy under an external electric field: $Q_i$ is the charge and $V_i$ is the electric potential on site $i$. The second term is the nearest-neighbor (NN) Coulomb repulsion. A is the Coulomb coefficient, which is taken as the energy unit 1. The ($Q_M$ - $Q_I$) is also taken as unit 1. The typical lattice size used in the simulation is $L_x$=$L_y$=40; which equates to domain sizes of 25 nm per site. Open and periodic boundary conditions are applied to the $y$ and $x$ sides respectively, since the gate voltage is applied along the $y$ direction. The electric field within the lattice is simplified to be uniform. The dynamics of phase separated lattices are simulated using the standard Markov chain Monte Carlo method, and their resistances are calculated by mapping these lattices to resistor networks.[13,36,37] Figure 3 shows resistance as a function of applied electric field for a temperature corresponding to the fluid phase separated regime in which a small number of newly seeded metallic domains coexist with insulating domains in a ratio of 20/80 and are not locked in the glass phase. The inset Monte Carlo snapshots show the system states across the range of applied fields.

At low fields, the metallic clusters are randomly distributed. Increasing the field draws the metallic regions together and, at a critical value, overcomes the Coulombic repulsion between clusters thereby creating a percolation channel that reduces the resistance. Decreasing the electric field below the critical value returns the system to a disordered, high resistance state. At lower temperatures corresponding to the glass phase, the metallic regions are not free to migrate so switching cannot be realized without inducing phase transitions which would change the transition characteristics. It is also found that the volume fraction of metallic elements affects the required switching field, where a higher metallic concentration requires a lower electric field. This result is experimentally confirmed by changing the applied magnetic field on the device. (See Supporting Information) This model gives a good qualitative match to experimental observations; however future studies using scanning probe magnetic or resistive imaging will be needed for direct confirmation.

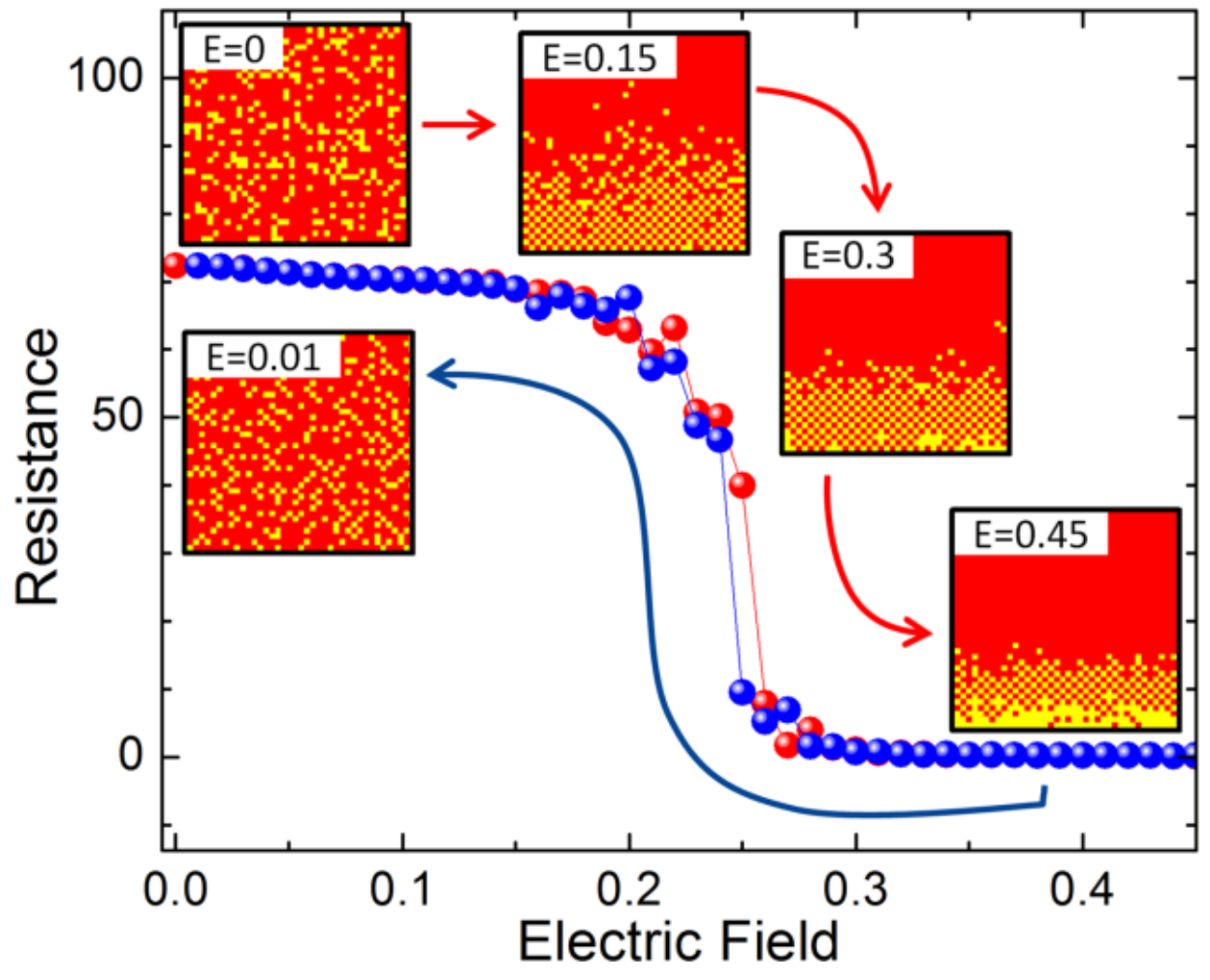

**Figure 3**. Electrophoretic model of resistive switching. Resistance vs applied field for model system of 40x40 matrix of 20% metal(yellow) and 80% insulating(red) elements above the critical entropic temperature with system snapshots along field cycle. Matrix is taken to lie directly between a single set of gates. The disordered metallic and insulating domains coexist in a random non-percolative network under no electric field. The application of electric field induces electrophoretic-like domain movement. Beyond the switching field, the Columbic repulsion between like domains is overcome and domains coalesce into a single low resistance channel which opens a percolation path and results in a large drop in resistance.

There are several benefits to this type of electrophoretic-like device. Compared to other devices, which can degrade with use and time caused by ionic migration or suffer from dielectric breakdown, this design allows contact free, reversible control of the electronic states which we anticipate will greatly increase the device lifetime.[8] Figure 4a shows the resistance of the wire as it is cycled 10,000 times between gate biases of 0.2 V and 70 V; while further testing will be explored, over this range the two resistance states are stable and do not decay over time. Figure 4b demonstrates consistent resistance levels in the binned values of all collected resistance states.

Also of practical importance, while one set of gates can produce two stable resistive states, the addition of more gates along the wire length allows access to more unique resistive states. Figure 4c demonstrates a four level resistive device that is created with the use of two sets of gates. Each level is a combination of on and off states triggered by energizing/de-energizing different combinations of the two gates at a bias of 65 V. The variation in local resistance change can be attributed to the wire region between each gate possessing a unique, random distribution of metallic domains where the relative change to the percolation channel's resistance after switching is different for each region depending on how much of a pre-switch path was inherently present.[13,24] As such, fabricating a distribution of gate sizes or increasing the number of gates should give access to even more resistive levels with a relation $2^n$ where n = number of gates.

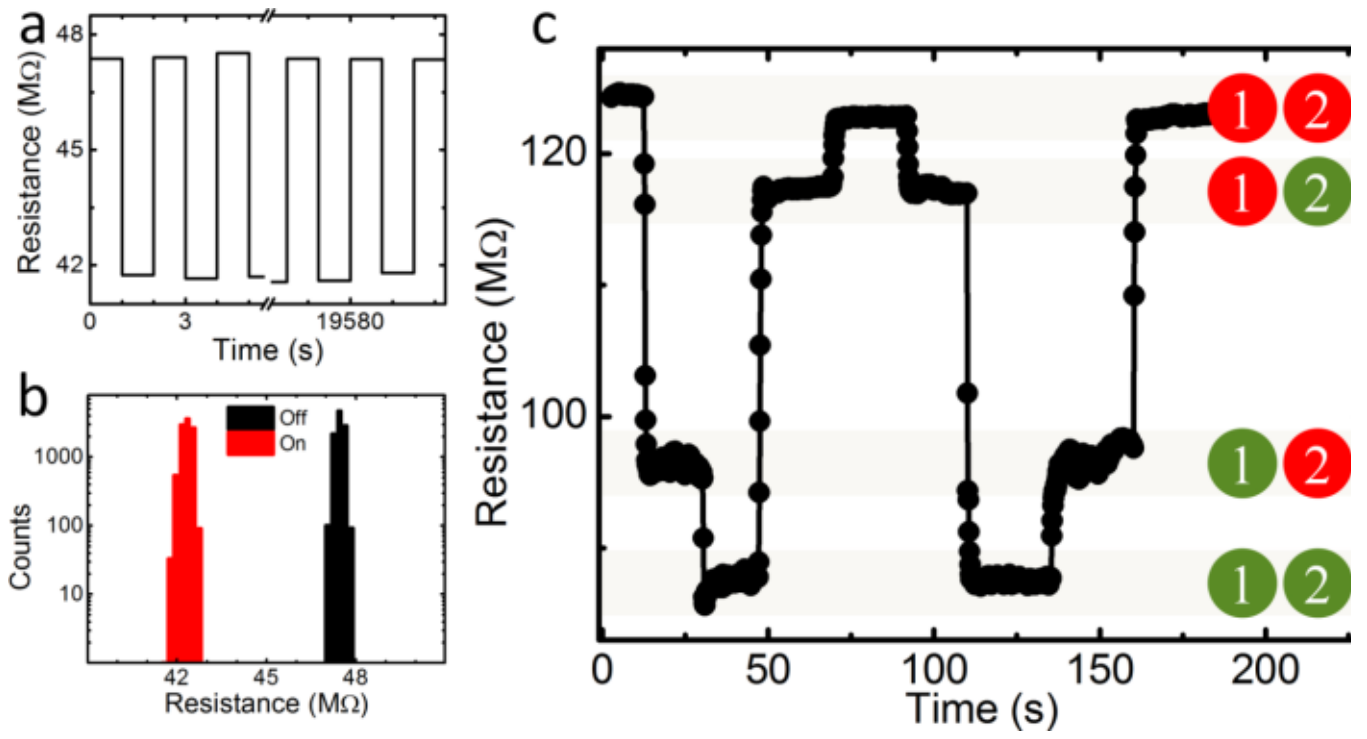

**Figure 4**. Effects of device cycling on 700 nm wire and example of multi-level switching. (a)Horizontal step plot of resistance vs time where a 70 V bias is applied and removed with a full cycle every 2 seconds. Resistance levels are nearly identical at the beginning and end of 10,000 cycles with no sign of aging. (b)Histogram of resistive levels collected from 10,000 voltage cycles show two clear resistive levels corresponding to the biased and unbiased states. (c)Four level resistive behavior achieved by controlling two pairs of gates simultaneously on the same 400nm wide wire. 1 corresponds to the central gate with 2 corresponding to a gate nearer the probe current source. Red represents that the gate voltage is off. The lowest resistance state corresponds to a 65 V bias applied to both gates while the highest resistance state is reached with no voltage applied to either gate.

Though these properties are highly desirable, the presented LPCMO prototype system is unlikely to find immediate application due to its active region being below room temperature, use of relatively high magnetic fields to adjust domain sizes, and slow switching speeds on the order of 300 ms. (See Supporting Information) However, these findings open the door to a wealth of other electronically phase separated materials exhibiting electronic phase separation across a wide spectrum of temperatures, domain sizes, and resistive and magnetic properties.[3–6,38,39] Indeed, one of the most exciting aspects of this new type of switching behavior is that it may be present across such a wide range of complex materials, from high temperature superconductors to multiferroics. Since many of these materials are sensitive to multiple types of fields, have a broad range of desirable characteristics, and offer unique blends of electronic phases, there are many opportunities to implement new types of functionalities which can be tailored to specific temperature and size specifications.

The prototype manganite device described in this work demonstrates a new type of switching mechanism that may be accessible in many complex oxide materials. Our phenomenological model suggests that the abrupt resistance changes observed upon application of a sufficiently high electric field is driven by electrophoretic-like redistribution of the coexisting electronic phase domains. The geometric control of electronic element configurations may act as rudimentary rewritable circuitry where nanoscale phenomena operating at mesoscale lengths can be used to drive macroscopic transport. Further development of the concepts presented may lead to practical low power switching applications, find use in reconfigurable interconnects in VLSI chips, and promises to bring new types of multifunctionality to a wide range of other complex materials where electronic phase competition exists.

**Supporting Information**. Sample preparation and characterization details, further transport and E-field model details, and switching lifetime discussion. This material is available free of charge via the Internet at http://pubs.acs.org.

**Corresponding Authors**

*E-mail: 5zw@ornl.gov and shenj5494@fudan.edu.cn.

ACKNOWLEDGMENT

This effort was supported by the US DOE, Office of Basic Energy Sciences, Materials Sciences and Engineering Division, (TZW, ED, and XX) and under US DOE grant DE-SC0002136 (HWG). Nanofabrication (PDR, JHN) and Magnetization measurements (ZG) were conducted at the Center for Nanophase Materials Sciences, which is sponsored at Oak Ridge National Laboratory by the Scientific User Facilities Division, Office of Basic Energy Sciences, U.S. Department of Energy JHN also acknowledge support from the Joint Institute of Advanced Materials Partial support was also supplied from the National Science Foundation of China No. 11274060 (SD), and the National Basic Research Program of China (973 Program) under grant No. 2011CB921801 (JS).

REFERENCES

(1) Waser, R.; Dittmann, R.; Staikov, G.; Szot, K. *Advanced Materials* **2009**, *21*, 2632–2663.

(2) Shenoy, V. B.; Sarma, D. D.; Rao, C. N. R. *ChemPhysChem* **2006**, *7*, 2053–2059.

(3) Sharoni, A.; Ramírez, J. G.; Schuller, I. K. *Physical Review Letters* **2008**, *101*, 26404.

(4) Basov, D. N.; Averitt, R. D.; Van der Marel, D.; Dressel, M.; Haule, K. *Reviews of Modern Physics* **2011**, *83*, 471–541.

(5) Mathur, N. D.; Littlewood, P. B. *Solid State Communications* **2001**, *119*, 271–280.

(6) Tokura, Y. *Colossal Magnetoresistive Oxides*; Gordon and Breach: Amsterdam, 2000.

(7) Wu, T.; Ogale, S. B.; Garrison, J. E.; Nagaraj, B.; Biswas, A.; Chen, Z.; Greene, R. L.; Ramesh, R.; Venkatesan, T.; Millis, A. J. *Physical Review Letters* **2001**, *86*, 5998–6001.

(8) Dhoot, A. S.; Israel, C.; Moya, X.; Mathur, N. D.; Friend, R. H. *Physical Review Letters* **2009**, *102*, 136402.

(9) Chen, Y.; Chen, L.; Lian, G.; Xiong, G. *Journal of Applied Physics* **2009**, *106*, 23707–23708.

(10) Sun, J.; Jia, C. H.; Li, G. Q.; Zhang, W. F. *Applied Physics Letters* **2012**, *101*, 133504–133506.

(11) Ward, T. Z.; Gai, Z.; Guo, H. W.; Yin, L. F.; Shen, J. *Physical Review B* **2011**, *83*, 125125.

(12) Xiang, P.-H.; Asanuma, S.; Yamada, H.; Inoue, I. H.; Sato, H.; Akoh, H.; Sawa, A.; Ueno, K.; Yuan, H.; Shimotani, H.; Kawasaki, M.; Iwasa, Y. *Advanced Materials* **2011**, *23*, 5822–5827.

(13) Liu, S. D. and C. Z. and Y. W. and F. Y. and K. F. W. and J.-M. *Journal of Physics: Condensed Matter* **2007**, *19*, 266202.

(14) Sacanell, J.; Parisi, F.; Campoy, J. C. P.; Ghivelder, L. *Physical Review B* **2006**, *73*, 14403.

(15) Ghivelder, L.; Freitas, R. S.; Das Virgens, M. G.; Continentino, M. A.; Martinho, H.; Granja, L.; Quintero, M.; Leyva, G.; Levy, P.; Parisi, F. *Physical Review B* **2004**, *69*, 214414.

(16) Dagotto, E.; Hotta, T.; Moreo, A. *Physics Reports* **2001**, *344*, 1–153.

(17) Asamitsu, A.; Tomioka, Y.; Kuwahara, H.; Tokura, Y. *Nature* **1997**, *388*, 50–52.

(18) Ahn, C. H.; Triscone, J.-M.; Mannhart, J. *Nature* **2003**, *424*, 1015–1018.

(19) Cao, J.; Wu, J. *Materials Science and Engineering: R: Reports* **2011**, *71*, 35–52.

(20) Cox, S.; Singleton, J.; McDonald, R. D.; Migliori, A.; Littlewood, P. B. *Nat Mater* **2008**, *7*, 25–30.

(21) Ward, T. Z.; Budai, J. D.; Gai, Z.; Tischler, J. Z.; Yin, L.; Shen, J. *Nat Phys* **2009**, *5*, 885–888.

(22) Jin, S.; Tiefel, T. H.; McCormack, M.; Fastnacht, R. A.; Ramesh, R.; Chen, L. H. *Science* **1994**, *264* , 413–415.

(23) Zhai, H.-Y.; Ma, J. X.; Gillaspie, D. T.; Zhang, X. G.; Ward, T. Z.; Plummer, E. W.; Shen, J. *Physical Review Letters* **2006**, *97*, 167201.

(24) Singh-Bhalla, G.; Selcuk, S.; Dhakal, T.; Biswas, A.; Hebard, A. F. *Physical Review Letters* **2009**, *102*, 77205.

(25) Wu, T.; Mitchell, J. F. *Physical Review B* **2006**, *74*, 214423.

(26) MurakamiY.; KasaiH.; J., K.; MamishinS.; ShindoD.; MoriS.; TonomuraA. *Nat Nano* **2010**, *5*, 37–41.

(27) Uehara, M.; Mori, S.; Chen, C. H.; Cheong, S.-W. *Nature* **1999**, *399*, 560–563.

(28) Sharma, P. A.; Kim, S. B.; Koo, T. Y.; Guha, S.; Cheong, S.-W. *Physical Review B* **2005**, *71*, 224416.

(29) Dhakal, T.; Tosado, J.; Biswas, A. *Physical Review B* **2007**, *75*, 92404.

(30) Ward, T. Z.; Liang, S.; Fuchigami, K.; Yin, L. F.; Dagotto, E.; Plummer, E. W.; Shen, J. *Physical Review Letters* **2008**, *100*, 247204.

(31) Milward, G. C.; Calderon, M. J.; Littlewood, P. B. *Nature* **2005**, *433*, 607–610.

(32) Müller, K. A.; Burkard, H. *Physical Review B* **1979**, *19*, 3593–3602.

(33) Freitas, R. S.; Mitchell, J. F.; Schiffer, P. *Physical Review B* **2005**, *72*, 144429.

(34) Garbarino, G.; Acha, C.; Levy, P.; Koo, T. Y.; Cheong, S.-W. *Physical Review B* **2006**, *74*, 100401.

(35) Wu, T.; Mitchell, J. F. *Applied Physics Letters* **2005**, *86*, 252503–252505.

(36) Dong, S.; Zhu, H.; Wu, X.; Liu, J.-M. *Applied Physics Letters* **2005**, *86*, 22501–22503.

(37) Mayr, M.; Moreo, A.; Vergés, J. A.; Arispe, J.; Feiguin, A.; Dagotto, E. *Physical Review Letters* **2001**, *86*, 135–138.

(38) Becker, T.; Streng, C.; Luo, Y.; Moshnyaga, V.; Damaschke, B.; Shannon, N.; Samwer, K. *Physical Review Letters* **2002**, *89*, 237203.

(39) Shenoy, V. B.; Rao, C. N. R. *Philosophical Transactions of the Royal Society A: Mathematical, Physical and Engineering Sciences* **2008**, *366* , 63–82.